\documentclass[aps,prb,floatfix,twocolumn,showpacs]{revtex4-1}
\usepackage{graphicx}
\usepackage{dcolumn}
\usepackage{bm}

\begin{document}

\title{Multiferroic behavior of Aurivillius Bi$_4$Mn$_3$O$_{12}$ from first-principles}

\author{Silvia Tinte}
\affiliation{INTEC-CONICET,
Universidad Nacional del Litoral,
G\"{u}emes 3450, (3000) Santa Fe, Argentina}

\author{M. G. Stachiotti}
\affiliation{IFIR-CONICET,
Universidad Nacional de Rosario, 
27 de Febrero 210 Bis (2000) Rosario, Argentina}

\date{\today}

\begin{abstract}
The multiferroic behavior of the hypothetical Aurivillius compound 
Bi$_4$Mn$_3$O$_{12}$ has been explored on the basis of density functional calculations.
We find that the tetragonal paraelectric phase of this material is 
ferromagnetic, showing ferroelectric and antiferrodistortive 
instabilities similar to the ones observed 
in its ferroelectric parent compound Bi$_4$Ti$_3$O$_{12}$. 
Our results indicate, however, that the presence of Mn$^{+4}$ ions 
at the $B$-sites shrinks the cell volume and consequently the 
unstable polar mode, associated with the ferroelectric polarization,
is overcame by an antiferrodistortive distortion. In this way, 
Bi$_4$Mn$_3$O$_{12}$ exhibits incipient ferroelectricity at its 
equilibrium volume. We show that the ferroelectric state 
can be favored by strain or partial substitution of Mn with Ti.  

\end{abstract}


\maketitle

\section{Introduction}

Multiferroic materials, exhibiting both ferroelectricity and magnetic order in 
the same phase, are of particular interest for their potential technological applications. 
One of the current trends in the search for new multiferroic compounds is, 
starting from a ferroelectric host, to incorporate magnetically active species 
and check whether it is (anti)ferromagnetic and insulating. An alternative 
way to predict multiferroic behavior is from material computational design, 
that is the prediction of multiferroic properties using first-principles techniques
prior to their experimental verification.~\cite{spaldin03,ederer05,picozzi09} 
Examples of this methodology include the prediction of a ferroelectric instability 
in BiMnO$_3$,~\cite{hill99} several unstable phonon branches in BiCrO$_3$,~\cite{hill02}
structural instabilities of CaMnO$_3$ \cite{bhatta09} and the properties of 
as-yet-unsynthesized 3$d$-5$d$ double perovskites.~\cite{lezaic11}         

Between the ferroelectric materials, Aurivillius layered perovskites are good trials
for multiferroics. 
Described by the formula [Bi$_2$O$_2$][A$_{n-1}$B$_n$O$_{3n+1}$], 
they are formed by stacking fluorite blocks of Bi$_2$O$_2$ with $n$ perovskitelike blocks.
Various $A$ and $B$ cations are allowed, so that this family has numerous representatives. 
Archetype compounds are  Bi$_2$WO$_6$ (n=1), SrBi$_2$Ta$_2$O$_9$ (n=2) and 
Bi$_4$Ti$_3$O$_{12}$ (n=3).
Recently some experimental groups have synthesized Aurivillius compounds 
incorporating magnetic ions at the $B$-site. The underlying idea was the
substitution of magnetic transition metal cations into the central 
octahedron layer of the perovskite-type block, while maintaining 
ferroelectric displacements in the outer octahedron layers. 
For that reason, those works involved high-order Aurivillius phases (n$>$3), 
such as Bi$_5$Ti$_3$FeO$_{15}$,~\cite{mao08,dong08,srinivas99}
Bi$_6$Ti$_3$Fe$_2$O$_{18}$,~\cite{srinivas99}
and Bi$_5$Ti$_3$CrO$_{15}$.~\cite{giddings11}
Sharma {\it et al.} reported ferroelectric behavior in a family of transition metal 
substituted three-layer Aurivillius phases, formulated
Bi$_{2-x}$Sr$_{2+x}$(Nb/Ta)$_{2+x}$M$_{1-x}$O$_{12}$ 
(x $\approx$ 0.5, M = Ru$^{4+}$, Ir$^{4+}$ or Mn$^{4+}$), showing that 
these compounds have the same orthorhombic symmetry and similar dielectric 
and ferroelectric properties as their (non-magnetic) ferroelectric parent 
compounds.~\cite{sharma08,sharma07} Several investigations have been based 
on the incorporation of perovskite like LaFeO$_3$, BiMnO$_3$, and BiFeO$_3$ 
into a Bi$_4$Ti$_3$O$_{12}$ matrix.~\cite{wu11,mao09,zurbuchen07} 
However, one can find that the reported magnetic properties in these 
materials are quite weak even at low temperature. The modification of 
Bi$_4$Ti$_3$O$_{12}$ ceramics was also investigated by substituting
their cations partly with Gd,~\cite{khomchemko10} Fe,~\cite{chen10}
and Mn.~\cite{ting10}

In spite of the large amount of experimental work showing that the generic 
structure of Aurivillius phases has the potential to serve as a template for 
multiferroic behavior, theoretical studies that help to clarify the effect of 
magnetic species in the Aurivillius structure are missing.
In this paper we investigate the effects of the substitution of Ti ions by 
magnetically active Mn$^{+4}$ ions at the $B$ sites of Bi$_4$Ti$_3$O$_{12}$ (BIT),  
which is a prototypical Aurivillius compound with bismuth also occupying the 
perovskite $A$ site. BIT has received tremendous attention during the last decade, 
especially in the thin film form, due to its promise in nonvolatile ferroelectric 
random access memory applications.~\cite{scott00} It has a reasonably large 
spontaneous polarization ($\sim$ 50 $\mu$C/cm$^2$), a high Curie temperature (950 K), 
and fatigue resistance on Pt electrodes. Its complex crystal structure can be 
described in terms of relatively small perturbations from a high-symmetry 
body-centered tetragonal structure (space group symmetry I4/mmm), which contains 
only one formula unit per primitive cell. Two main distortions from this tetragonal 
(paraelectric) structure lead to the ferroelectric orthorhombic phase. First, the ions 
displace along the [110] axis of the tetragonal structure. Second, the TiO$_6$ octahedral
rotate around the $a$ and $c$ axes.~\cite{rae90} The first distortion is directly 
responsible for the observed macroscopic polarization along the orthorhombic $a$ direction 
([110] axis of the tetragonal structure).

Specifically, we performed density-functional theory calculations 
to investigate the hypothetical three-layer Aurivillius compound 
Bi$_4$Mn$_3$O$_{12}$ (BIM), which results from substituting the three 
$B$-site Ti$^{4+}$ cations with Mn$^{4+}$ in the BIT lattice. 
We predict structural, electronic and magnetic properties of BIM, 
deepening the understanding of the effects of magnetic species in the Aurivillius
structure of BIT. The paper is organized as follows. In Sec. II we describe 
computational details. Results are presented in Sec. III. Within this section we 
compare our BIM results with those of BIT from previous works and also with results
we have obtained. First, the tetragonal structure of BIM is optimized and the 
electronic structure and magnetic order of this phase are discussed. Then we focus 
on the two more unstable modes of BIT: the zone-center $E_u$ modes related 
to the polar instability and the zone-boundary $X{_3^+}$ modes related to 
oxygen antiferrodistortive (AFD) distortions. 
We inspect then the energy landscape along the unstable modes directions and 
find that by replacing the Ti cations with Mn, the polar $E_u$ mode is overcome 
by the $X{_3^+}$ mode. Finally, we investigate the possibility of favoring the 
ferroelectric distortion by strain and chemical engineering.

\section{Computational details}
We performed standard first-principles calculations using density-functional theory 
within the generalized gradient approximation (GGA) with the Perdew, 
Burke and Ernzerhof~\cite{PBE} parametrization as implemented in 
the VASP-4.6 code.~\cite{vasp} The projector augmented wave potentials 
were used including explicitly 15 valence electrons for Bi (5$d^{10}$ 6$s^2$ 6$p^3$), 
13 for Mn (3$p^6$, 3$d^5$, 4$s^2$), 10 for Ti (2$p^6$, 3$d^3$, 4$s^1$) 
and 6 for oxygen (2$s^2$, 2$p^4$). Convergence with respect to the $k$-point mesh 
was achieved using a 6$\times$6$\times$6 $\Gamma$-centered grid.
For a proper treatment of strongly localized Mn $d$ electrons,  
we used the on-site Coulomb correction (GGA+$U$) in the Dudarev 
implementation with parameter $U_{eff}=U-J=5.2~eV$.~\cite{noteU}
Except for the calculations done to determine the magnetic ground-state, 
where different U$_{eff}$ values were used to check the energy differences 
between spin configurations, all other calculations were performed using 
the mentioned value.

Using a 19-atom body-centered-tetragonal unit cell with space group I4/mmm,
we relax all degrees of freedom (lattice parameters and internal coordinates  
along the stacking direction $\it z$) for spin polarized structures.  
We optimized the ground state structures with a plane-wave cutoff of 600~eV
which yielded good convergence and, hence, was also used for the rest of the 
calculations. To search for the presence of ferroelectric and antiferrodistortive 
instabilities in the tetragonal paraelectric structure, we determined the phonon 
frequencies and eigenvectors of the infrared-active $E_u$ and rotational 
$X{_3^+}$ modes, which are the main unstable modes of BIT.
To this end, we calculated atomic forces of symmetry-adapted configurations  
determined with the help of the ISOTROPY software,\cite{isotro}
where selected atoms are displaced by 0.01~\AA~along the $x$, $y$ and/or $z$ 
direction. To compute the $X{_3^+}$ modes, in particular, the primitive cell 
was doubled to a 38-atom conventional tetragonal cell. From the forces as a 
function of displacements, a dynamical matrix of 11$\times$11 (8$\times$8) 
for $E_u$ ($X{_3^+}$) symmetry was constructed and diagonalized. 
Once the eigenvectors have been determined, we evaluate the total energy as 
a function of the displacement pattern of the unstable (imaginary frequency) modes. 
Those curves will provide the ferroelectric and antiferrodistortive instability 
energy associated with a particular phonon mode.

\section{Results}
We begin by describing the optimized tetragonal structure of BIM.
The minimized lattice parameters are a = 3.811~\AA~and c = 32.613~\AA.
These values are smaller than those we compute for BIT using 
GGA: a = 3.842~\AA~and c = 33.241~\AA, which in turn differ slightly from 
the experimental values reported in the literature (a = 3.8633~\AA~and 
c = 33.2942~\AA,\cite{herv99} and a = 3.850~\AA~and c = 32.832~\AA)\cite{rae90}.
As seen the replacement of Ti$^{4+}$ ions with Mn$^{4+}$ shrinks the equilibrium volume 
of the unit cell, which is consistent with the fact that six coordinated 
Mn$^{4+}$ has an ionic radius (0.67~\AA) smaller than that of Ti$^{4+}$ (0.745~\AA).
This chemical substitution would be analogous to applying pressure on BIT and 
one could expect, following the general trend in many perovskite oxides, 
a strong sensitivity of the ferroelectric instability to volume. 

\begin{figure}
\begin{center}
\includegraphics[width=5.cm,angle=0]{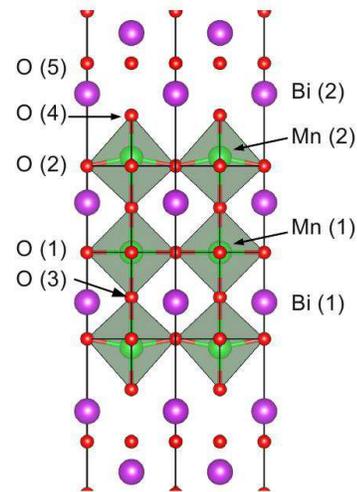}
\end{center}
\caption{(Color online) BIM tetragonal phase along [001] from 0 to $c$/2.}
\label{fig:auriv-struct}
\end{figure}

In the perovskitelike blocks of the tetragonal structure, by symmetry there is one 
central Mn$^{4+}$ ion [Mn(1) in Fig.~\ref{fig:auriv-struct}] and two equivalent 
Mn$^{4+}$ ions above and below the central layer, Mn(2). In the optimized structure 
of BIM, the two Mn(2) cations are pushed away along the $c$-axis from the center 
of each oxygen octahedron towards the fluorite blocks. Besides, the oxygen octahedral 
around these Mn(2) ions are considerably elongated along the $c$-axis, opposite 
to the undistorted central octahedron. 
The $A$-site Bi$^{3+}$ ions [Bi(1)] also move towards the fluorite blocks, that is, 
they approach the O(2) ions, moving away from the O(1) ions. 
Table~\ref{tab:posi} lists the optimized internal atomic positions of BIM.
For comparison, we include our GGA results for BIT and the respective experimental 
values from Ref.~\onlinecite{herv99}. As seen the internal coordinates in both 
compounds are similar.

\begin{table}
\caption {Wyckoff positions for relaxed BIM tetragonal structure are compared 
to GGA calculations and experimental results for BIT.}
\begin{ruledtabular}
\begin{tabular}{lcccc}
Atom & Site & BIM & BIT & BIT exp.\cite{herv99}    \\
\hline
Bi1 &	e &   0.4342  &  0.4336 & 0.4325   \\ 
Bi2 &	e &   0.2901  &  0.2887 & 0.2885   \\ 
Mn1 &	a &   0.0000  &  0.0000  & 0.0000    \\ 
Mn2 &	e &   -0.1254 & -0.1275 & -0.1293 \\ 
O1 &	c &   0.00000   &  0.0000  & 0.0000   \\ 
O2 &	g &   0.3858  &  0.3848 & 0.3823 \\ 
O3 &	e &   -0.0593 & -0.0595 & -0.0589  \\ 
O4 &	e &   -0.1810 & -0.1815 & -0.1817  \\ 
O5 &	d &    0.0000  &  0.0000  & 0.0000   \\
\end{tabular}
\end{ruledtabular}
\label{tab:posi}
\end{table}

The magnetic order of BIM was investigated by inspecting the total energy 
of different configurations with spin pointing along [001]. We have optimized 
separately the tetragonal structure for the ferromagnetic (FM) ordering and 
for the following antiferromagnetic (AFM) arrangements in the perovskitelike 
block: $A$, $C$, and $G$-type, in which the fraction of parallel aligned 
first-neighbor spins is 2/3, 1/3, and 0, respectively. By comparing the total 
energy of those configurations we find that the magnetic ground state of 
the material is FM with local magnetic moments of 3.3 and 3.4~${\mu_B}$ for 
Mn(1) and Mn(2), respectively, which are slightly higher than the formal value 
of 3~${\mu_B}$ expected for a four-valence Mn ($d^{3}$) atom.
We find that the $A$-type AFM arrangement is located at an energy slightly higher
than the FM state, just a few mRy/f.u.~above it. The energy of the $C$-type AFM
ordering is more separated, $\sim$ 15~mRy/f.u. above the FM state. Finally, the 
$G$-type AFM  configuration lies a few mRy/f.u. above the $C$-type.
It is known that the energy differences between spin configurations are directly 
related to the magnetic interactions or exchange constants, which depend on 
the on-site Coulomb repulsion parameter U$_{eff}$. Then, we have verified that 
the FM order is the most stable configuration for different choices of U$_{eff}$.
Although the variation of U$_{eff}$ produced quantitative changes, the hierarchy 
of the magnetic configurations is kept. We point out that the optimized crystal 
structure of BIM described above corresponds to the FM ground state. Furthermore, 
all of the following results have also been obtained using the FM spin configuration.

\begin{figure}
\begin{center}
\includegraphics[width=7.8cm,angle=0]{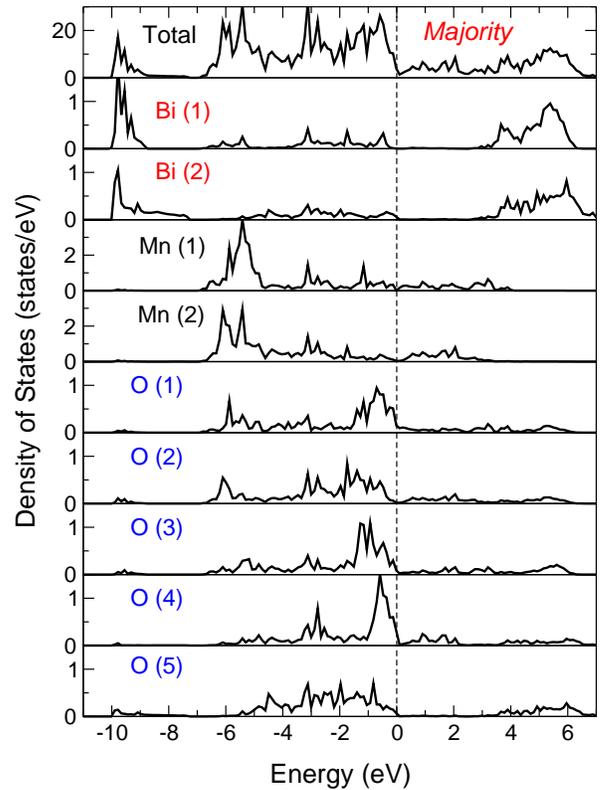}
\end{center}
\caption{Total majority-spin state and projected DOS onto atomic sites 
for tetragonal BIM. Zero energy is set as the Fermi level.}
\label{fig:dos_BIM}
\end{figure}

\begin{figure}
\begin{center}
\includegraphics[width=8.cm,angle=0]{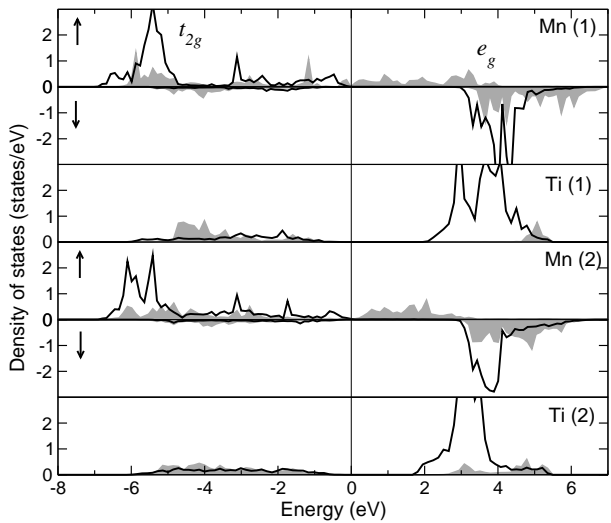}
\end{center}
\caption{DOS projected onto $t_{2g}$ (black curves) and $e_g$ (gray dashed regions) orbitals 
for each Mn site in BIM for the majority and minority spins. They are compared to the respective 
orbitals of the Ti cations in BIT. The DOS of BIT was computed at the 
BIM equilibrium volume.}
\label{fig:dos_Mn}
\end{figure}

The total density of states (DOS) and projected DOS onto the atomic sites for 
the majority spins are shown in Fig.~\ref{fig:dos_BIM}. The system has no band 
gap in the paraelectric I4/mmm phase, similar to what was observed for the 
paraelectric phase of other $A$MnO$_3$ perovskites.~\cite{hill99,li11}
Instead, a half-metallic electronic structure is achieved with a band gap of 
$\sim$ 2.2~eV in the minority spins (not shown here). The occupied states 
of the conduction band (ranging from -7~eV to the Fermi level) are mainly O~$2p$ and 
Mn~$3d$ states, which are strongly hybridized. The low-energy Bi states, 
around -10~eV, come from the $6s$ orbitals (the so-called lone pairs), 
while the unoccupied Bi~$6p$ states are located at high energies around 6 eV. 
The contribution of the Bi $6s$ and $6p$ states to the conduction band indicates 
that these states are hybridized with the $2p$ of their neighbor oxygens, 
which is similar to what has been reported for BIT.\cite{machado04}   
To better visualize the differences between BIM and BIT,  
the partial DOS of the $B$-site ion which is in an octahedron crystal field,
are projected onto the $t_{2g}$ and $e_{g}$ orbitals (black lines
and gray dashed areas in Fig.~\ref{fig:dos_Mn}, respectively).
The DOS of BIT was obtained at the BIM equilibrium crystal structure 
to avoid differences that could arise from the different lattice 
parameters of the two compounds. It is clear from the figure that 
the majority occupied Mn~$3d$ states are much more extended than those of Ti, 
showing that the Mn~$3d$ hybridize more strongly with the O~$2p$ states.
The majority-spin Mn $t_{2g}$ states are fully occupied whereas the corresponding 
minority spins are practically empty. Concerning the $e_{g}$ orbitals, the 
majority spins are more extended and they cross the Fermi level, 
which is responsible for the paraelectric phase being half-metallic.

Next we investigate the possibility of ferroelectricity in BIM by studying 
the unstable normal modes of the high-symmetry I4/mmm tetragonal phase, and 
compare our results with those for BIT. We note that as BIM is a hypothetical 
compound, we do not have an experimental low-temperature ground-state structure to 
decompose and identify the symmetry of active modes that would freeze in the 
high-temperature tetragonal phase. That limitation leads us to analyze the 
normal modes in tetragonal BIM with the same symmetry of the relevant modes 
found in the tetragonal parent compound BIT. Furthermore we are aware that 
the energy landscape of Aurivillius compounds, as it was shown in BIT~\cite{perez08} 
and SrBi$_2$Ta$_2$O$_9$,~\cite{perez04}
is much more complex than that of typical ferroelectric oxides, in which 
the freezing of one unstable polar mode directly explains the ferroelectric 
phase. So, based on an exhaustive study of the BIT phonon modes by 
Perez-Mato $\it et.~al$,~\cite{perez08} we restrict our discussion to zone 
center polar modes with $E_u$ symmetry and zone boundary antiferrodistortive 
modes with $X{_3^+}$ symmetry. We find that two of the 11 bi-dimensional 
$E_u$ modes of BIM have imaginary frequency, similarly to what was obtained 
for BIT.~\cite{machado04, perez08} The most unstable $E_u$ mode 
[$E_u$(1), $\omega= i146~cm^{-1}$] involves primarily antiphase displacements 
of the Bi(1) ions with respect to the apical oxygen O(3), plus a smaller 
displacement of the central Mn(1) against its neighbor oxygens O(1).
A similar pattern was indeed associated with the in-plane ferroelectric 
polarization of BIT.\cite{rae90}
The second unstable mode [$E_u$(2), $\omega= i57~cm^{-1}$] can be roughly 
described by the displacements of the Bi$_2$O$_2$ layers relative to the 
perovskitelike blocks. This kind of instability was found in other Aurivillius family
members,~\cite{stachiotti00} and seems to be a general feature of them.
Surprisingly, not only the eigenvectors but also the frequencies of the two 
unstable $E_u$ modes are very similar to the ones obtained in a 
linearized augmented plane-wave (LAPW) calculation of 
BIT, where $\omega= i145~cm^{-1}$ and $\omega= i48~cm^{-1}$ were assigned for the
$E_u$(1) and $E_u$(2) modes respectively.~\cite{machado04}
Regarding the zone-boundary modes, the most unstable mode [$X{_3^+}$(1), 
$\omega= i242~cm^{-1}$], is much more negative than the polar $E_u$(1) mode.
This mode takes place mainly at the center of perovskite slabs, tilting 
the central octahedron around the [110] direction, $a^{-}a^{-}c^{0}$ in the 
Glazer notation. However, as the outer apical O(4) oxygens almost do not move,
the two external octahedral are distorted. The second unstable AFD mode 
[$X{_3^+}$(2)] consists mostly of the tilting of the two external octahedral 
around the [110] direction. This mode involves the displacement of 
the O(4) oxygens as well as the Bi(2) ions belonging to the fluorite layer.
In summary, the described eigenvectors of the four unstable $E_u$ and $X{_3^+}$
modes are very similar to those obtained in BIT (illustrations of the 
different unstable eigenvectors of BIT are included in Ref.~\onlinecite{perez08}).

\begin{figure}
\begin{center}
\includegraphics[width=8.5cm,angle=0]{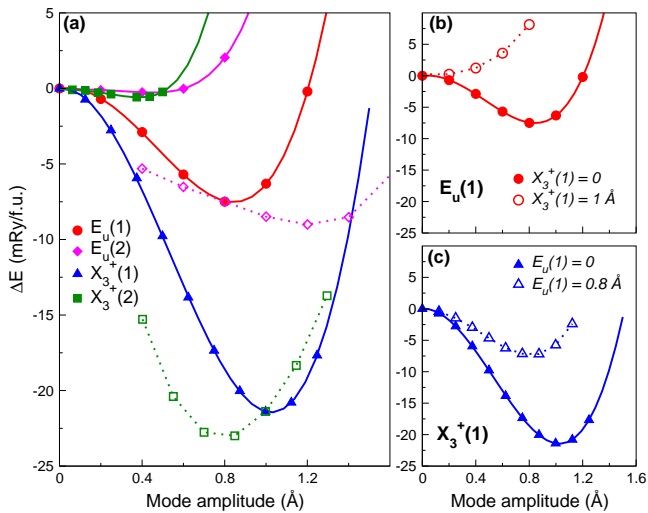}
\end{center}
\caption{(Color online) (a) Total energy per formula unit relative to the 
optimized tetragonal structure 
as a function of the unstable mode amplitudes with symmetry: $X{_3^+}$
(filled triangles and squares) and $E_u$ (filled circles and diamonds).
Combination of same symmetry modes: 
less unstable mode $X{_3^+}$(2) (open squares) and $E_u$(2) mode (open diamonds) 
are superimposed to the energy minima of the most unstable mode, $X{_3^+}$(1) and
($E_u$(1)), respectively. 
(b), (c) Coupling effect between the two most unstable modes 
of different symmetry.
(b) Change of energy per f.u.~as a function of the
$E_u$(1) mode amplitude relative to a structure with 
mode $X{_3^+}$(1) frozen with amplitude of 1~\AA~(open circles).
The curve for the pure $E_u$(1) distortion (filled circles) 
is repeated for comparison with. 
(c) Change of energy per f.u.~as a function of the
$X{_3^+}$(1) mode amplitude relative to a structure with  
mode $E_u$(1) frozen with amplitude of 0.8~\AA (open triangles),
and for the pure $X{_3^+}$(1) distortion (filled triangles).}
\label{fig:ene_BIM}
\end{figure}

The energy landscape around the tetragonal structure as a function of the 
amplitude in \AA~of the unstable modes is shown in Fig.~\ref{fig:ene_BIM}(a).
For each mode, the eigendisplacement vector is normalized with respect to the 
tetragonal high-symmetry 38-atom unit cell.
We summarize our results as follows.
First, the most unstable $X{_3^+}$(1) mode decreases the total energy in 
$\sim$ 21.4~mRy/f.u., against the $\sim$ 7.5~mRy/f.u. produced by the $E_u$(1) 
polar mode. This is strikingly different from what is observed in BIT where 
the rotational mode is overcame by the in-plane polar instability.
With regard to the other two modes, $X{_3^+}$(2) and $E_u$(2), 
they present a rather small energy instability. However, when each one by 
separate is combined with the most unstable mode of the same symmetry
(open symbols), they further decrease the energy minimum of the double-well 
obtained with the most unstable mode. 
Interestingly, the particular an-harmonic coupling of $X{_3^+}$(2) with $X{_3^+}$(1) 
(open squares) yields an absolute energy minimum.
The combination of these two rotational modes results in an atomic structure 
that differs from the high-symmetry phase mainly in that the 
central octahedron is slightly distorted and rotated around the 
[110] direction at an angle of $\sim$~10$^o$.
This structure, however, is still half-metallic and ferromagnetic, 
as is the high-symmetry one.
Finally, in Fig.~4(b) and 4(c), we explore the an-harmonic coupling between the 
two most unstable modes of different symmetry, $X{_3^+}$(1) and $E_u$(1), 
to investigate whether they cooperate in the energy minimization.
We find that the AFD $X{_3^+}$(1) distortion fully stabilizes 
the $E_u$(1) polar mode [open circles in Fig.~4(b)],
when the last one is computed from the structure corresponding to 
the energy minimum of $X{_3^+}$(1), compared to the pure $E_u$(1) distortion
from the optimized tetragonal structure.
Instead, the polar mode effect on the AFD $X{_3^+}$(1) [open triangles in panel (c)] 
is not so strong, but it does reduce the AFD double-well instability.
Therefore, these behaviors indicate that the coupling of $X{_3^+}$(1) and $E_u$(1) 
is penalized.

\begin{table}  
\caption {BIM relaxed lattice parameter in the tetragonal structure at   
two different in-plane strains and in the equilibrium value.}  
\begin{ruledtabular} 
\begin{tabular}{lcr} 
In-plane strain & a (\AA) &   c  (\AA)   \\ 
\hline  
-2~\%  &   3.7345 &  33.5498   \\   
0   &  3.8108 &  32.6128  \\   
+2~\%  &   3.8870 &  32.0985  \\  
\end{tabular}  
\end{ruledtabular}  
\label{tab:st}  
\end{table}  
  
Based on the existence of the polar instability in tetragonal BIM   
(although it is overcame by the rotational distortion), we investigate to 
what extent it would be possible to favor ferroelectricity by the application 
of epitaxial strain. We analyze two different structures around the 
equilibrium volume with in-plane lattice parameter, $a_{+2\%}$ and $a_{-2\%}$. 
The lattice parameters obtained by fixing the in-plane lattice constants, and 
fully relaxing $c$ and the atomic positions, are listed in Table~\ref{tab:st}. 

\begin{figure}
\begin{center}
\includegraphics[width=8.cm,angle=0]{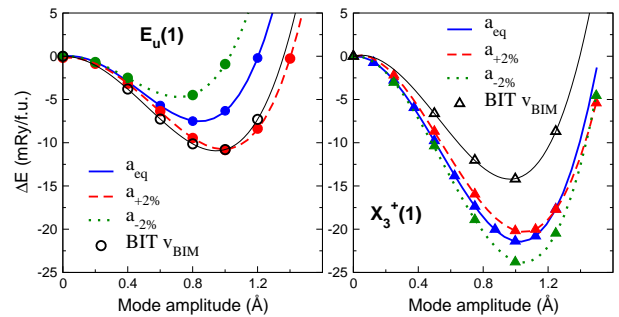}
\end{center}
\caption{(Color online) Effect of epitaxial strain on the two most unstable 
normal modes: $E_u$ (left panel) and $X{_3^+}$ (right panel). Energy differences 
relative to the optimized tetragonal structure for $a_{eq}$ (filled lines),
$a_{+2\%}$ (dashed lines), and $a_{-2\%}$ (dotted lines) using the eigenvectors 
obtained at the equilibrium volume. Open symbols correspond to BIT at the 
BIM equilibrium volume.}
\label{fig:ene_BIMstr}
\end{figure}

Figure \ref{fig:ene_BIMstr} displays the energy landscape of the strained 
structures and compares them with the stress-free one.
As expected, the in-plane $E_u$ mode (left panel) is favored by tensile 
bi-axial strain, whereas the AFD $X{_3^+}$ instability (right panel) is disfavored. 
This behavior, which is typical in $AB$O$_3$ perovskite oxides, indicates that 
epitaxial strain seems to be an appropriate route to induce ferroelectricity 
in this Aurivillius compound. However, it can be estimated by extrapolating 
the above data that it is necessary to apply a strain of $\sim +6\%$ to favor 
the polar mode over the antiferrodistortive distortion. This crudely estimated 
value is out of the range of strains usually accessible through epitaxial growth.

Since applying a large epitaxial strain is not easily achievable in practice, 
we instead explore the possibility of making BIM ferroelectric under a partial 
substitution of Mn by Ti. As a first test, we take the BIM equilibrium structure 
and replace all of the Mn ions with Ti; that is, we analyze BIT at the BIM equilibrium volume.
The resulting energy curves are also plotted in Fig.~\ref{fig:ene_BIMstr} 
with open symbols. By comparing them with the BIM curves (filled lines), 
we see that effectively the presence of the Ti ions favors the polar 
instability and considerably decreases the AFD one.
We can then speculate that a partial substitution of Mn with Ti would also favor 
ferroelectricity. To support this idea we check the energetics of the Aurivillius 
compounds Bi$_4$(TiMn$_2$)O$_{12}$ with the $B$ sites in the perovskitelike layer 
occupied by one Ti at the central octahedron, and two Mn ions in the external ones.
We set the lattice parameters $a$ and $c$ by interpolating those values between 
the optimized BIM and BIT ones. 
We find that the polar mode $E_u$ indeed is more unstable than the AFD one 
by $\sim$ 3.8~mRy/f.u., which shows that the Ti ion at the central octahedron 
plays an important role in stabilizing the ferroelectric mode over the AFD one. 
This is consistent with the fact that the two modes of interest, 
as described above, are localized in the center of the perovskite slab.
We think that this particular solid solution deserves a more detailed study in 
the future. Nevertheless, we can conclude that our results indicate that, 
not only the cell volume, but also the nature of the transition metal ion is 
indeed relevant for the stabilization of the ferroelectric phase in Aurivillius 
compounds. Therefore, it would be also interesting to further investigate the effect of 
magnetic species with higher ionic radius than Mn$^{4+}$, such as Ru$^{4+}$ and V$^{4+}$. 

In summary, we have investigated from first-principles the effects of 
magnetic Mn species in the Aurivillius structure of Bi$_4$Ti$_3$O$_{12}$. 
Specifically we explored the possibility of finding multiferroic behavior in
the hypothetical Aurivillius compound Bi$_4$Mn$_3$O$_{12}$, which results from 
replacing the Ti with Mn ions. We have found that the paraelectric tetragonal 
structure of this material presents ferromagnetic order and structural 
instabilities similar to BIT. We have shown, however, that the presence of Mn 
ions shrinks the cell volume, and the polar mode associated with the in-plane 
ferroelectric polarization, is overcame by an antiferrodistortive distortion. 
In this way, Bi$_4$Mn$_3$O$_{12}$ exhibits incipient ferroelectricity at its 
equilibrium volume. We have seen that although it is possible to favor the 
polar state by the application of epitaxial strain, the necessary strain value
is not easily achievable in practice. Our results indicate that 
the partial substitution of Mn with Ti is a fruitful alternative to favor 
the ferroelectric distortion, highlighting the relevant role played by the 
nature of the transition metal ion. We thus hope that our findings may be 
useful to better understand the multiferroic properties of Aurivillius
compounds.
 
This work was supported by  Grants No.~PICT-PRH99 and No.~PICT-2008(1867) from the ANPCyT-Argentina.
M.G.S. gives thanks for financial ssupport from Consejo de Investigaciones de la Universidad Nacional de Rosario.


\begin{thebibliography}{999}
 
\bibitem{spaldin03}
N. A. Spaldin and W. E. Pickett,
J. Solid State Chem. {\bf 176}, 615 (2003).

\bibitem{ederer05}
C. Ederer and N. A. Spaldin, 
Curr. Opin. Solid State Mater. Sci. {\bf 9}, 128 (2005).

\bibitem{picozzi09}
S. Picozzi and C. Ederer, 
J. Phys.: Condens. Matter {\bf 21} 303201 (2009). 
 
\bibitem{hill99}
N. A. Hill and K. M. Rabe, 
Phys. Rev. B {\bf 59}, 8759 (1999).

\bibitem{hill02} 
N. A. Hill, P. Baettig, and C. Daul,  
J. Phys. Chem. B {\bf 106}, 3383 (2002).  

\bibitem{bhatta09}
S. Bhattacharjee, E. Bousquet, and P. Ghosez 
Phys. Rev. Lett. {\bf 102}, 117602 (2009).

\bibitem{lezaic11}  
M. Lezaic and N. A. Spaldin,   
Phys. Rev. B {\bf 83}, 024410 (2011).  
  
\bibitem{mao08}  
X. Y. Mao, W. Wang, and X. Chen,   
Solid State Commun. {\bf 147}, 186 (2008).  
  
\bibitem{dong08}  
X. W. Dong, K. F. Wang, J. G. Wan, J. S. Zhu, J. Liu, 
J. Appl. Phys. {\bf 103}, 094101 (2008).  

\bibitem{srinivas99}  
A. Srinivas, S. V. Suryanarayana, G. S. Kumar and M. Mahesh Kumar,  
J. Phys.: Condens. Matter {\bf 11} 3335 (1999).   

\bibitem{giddings11}  
A. T. Giddings, M. C. Stennett, D. P. Reid, E. E. McCabe, C. Greaves, N. C. Hyatt, 
J. Solid State Chem. {\bf 184}, 252 (2011).  
  
\bibitem{sharma08}  
N. Sharma, B. J. Kennedy, M. M. Elcombe, Y. Liu and C. D. Ling,  
J. Phys.: Condens. Matter {\bf 20} 025215 (2008).   

\bibitem{sharma07}  
N. Sharma, C. D. Ling, G. E. Wrighter, P. Y. Chena, B. J. Kennedya, and  P. L. Lee,  
J. Solid State Chem. {\bf 180}, 370 (2007).  

\bibitem{wu11}  
Fei-Xiang Wu, Zhong Chen, Y. B. Chen, Shan-Tao Zhang, Jian Zhou,  
Yong-Yuan Zhu, and Yan-Feng Chen,   
Appl. Phys. Lett. {\bf 98}, 212501 (2011).  
  
\bibitem{mao09}  
X. Y. Mao, W. Wang, X. B. Chen, and Y. L. Lu,      
Appl. Phys. Lett. {\bf 95}, 082901 (2009).    

\bibitem{zurbuchen07}  
M. A. Zurbuchen, R. S. Freitas, M. J. Wilson, P. Schiffer, M. Roeckerath, J. Schubert, M. D. Biegalski  
G. H. Mehta, D. J. Comstock, J. H. Lee, Y. Jia, and D. G. Schlom,  
Appl. Phys. Lett. {\bf 91}, 033113 (2007).  

\bibitem{khomchemko10}  
V. A. Khomchenko, G. N. Kakazei, Y. G. Pogorelov, J. P. Araujo, M. V. Bushinsky, D. A. Kiselev,  
A. L. Kholkin, and J. A. Paixão,  
Mat. Lett. {\bf 64}, 1066 (2010).  
  
\bibitem{chen10}  
X. Q. Chen, F. J. Yang, W. Q. Cao, H. Wang, C. P. Yang, D. Y. Wang, K. Chen,
Solid State Comm. {\bf 150}, 1221 (2010).

\bibitem{ting10}  
J. Ting and B. J. Kennedy,   
J. Phys.: Conf. Ser. {\bf 251}, 012029 (2010).   
  
\bibitem{scott00}  
J. F. Scott, {\it Ferroelectric Memories}, Springer Series in Advanced  
Microelectronics, Vol. 3 (Springer-Verlag, Berlin, 2000).  
  
\bibitem{rae90}  
A. D. Rae, J. G. Thompson, R. L. Withers, and A. C. Willis,    
Acta Cryst., Sect. B: Struct. Sci. {\bf 46}, 474 (1990).  
  
\bibitem{machado04}  
R. Machado, M. G. Stachiotti, R. L. Migoni, and A. H. Tera,   
Phys. Rev. B {\bf 70}, 214112 (2004). 
  
\bibitem{perez08}  
J. M. Perez-Mato, P. Blaha, K. Schwarz, M. Aroyo, D. Orobengoa, I. Etxebarria, and A. Garcia,  
Phys. Rev. B {\bf 77}, 184104  (2008).  

\bibitem{perez04} 
J. M. Perez-Mato, M. Aroyo, and Alberto Garcia, P. Blaha, K. Schwarz, and J. Schweifer,
K. Parlinski, Phys. Rev. B {\bf 70}, 214111 (2004).
  
\bibitem{PBE}  
J. P. Perdew, K. Burke, and M. Ernzerhof,   
Phys. Rev. Lett. {\bf 77}, 3865 (1996). 
  
\bibitem{vasp}  
G. Kresse and J. Hafner, Phys. Rev. B {\bf 47}, R558 (1993).  
G. Kresse and J. Furthmuller, Phys. Rev. B {\bf 54}, 11169 (1996).  

\bibitem{noteU}
Note that as our hypotetical material prevents us from extracting U$_{eff}$ 
from experiments, we have taken that value from the literature for 
BiMnO$_3$ [Pio Baettig, Ram Seshadri, and Nicola A. Spaldin,
J. Am. Chem. Soc. {\bf 129}, 9854 (2007)] since this oxide has a similar 
structure to the perovskitelike blocks of the Aurivillius compound.
  
\bibitem{isotro}  
H. T. Stokes, D. M. Hatch, and B. J. Campbell, computer code  
ISOTROPY (stokes.byu.edu/isotropy.html).  
  
\bibitem{herv99}  
C. H. Hervoches and P. Lightfoot,   
Chem. Mater. {\bf 11}, 3359-3364 (1999).  
  
\bibitem{li11}  
N. Li, K.L. Yao, Z.Y. Sun, L.Zhu and G.Y. Gao  
J. Appl. Phys. {\bf 109}, 083715 (2011).  
  
\bibitem{stachiotti00}
M. G. Stachiotti, C. O. Rodriguez, C. Ambrosch-Draxl, and N. E. Christensen, 
Phys. Rev. B {\bf 61}, 14434 (2000).

\end{thebibliography}
\end{document}